\documentclass[a4paper,fleqn,usenatbib]{mnras}

\usepackage[T1]{fontenc}
\usepackage{ae,aecompl}

\usepackage{graphicx}	
\usepackage{amsmath}	
\usepackage{amssymb}	
\usepackage{multicol}        
\usepackage{bm}		
\usepackage{color}
\usepackage{xcolor}

\title[The hot limit of solar-like oscillations]
{The hot limit of solar-like oscillations from {\it Kepler} photometry}

\author[L. A. Balona]{\
L. A. Balona \thanks{E-mail: lab@saao.ac.za}
\\
South African Astronomical Observatory, 
P.O. Box 9, Observatory 7935, South Africa}

\begin{document}

\date{Accepted .... Received ...}

\pagerange{\pageref{firstpage}--\pageref{lastpage}} \pubyear{2011}

\maketitle

\label{firstpage}

\begin{abstract}
{\it Kepler} short-cadence photometry of 2347 stars with effective
temperatures in the range 6000 -- 10000\,K was used to search for the 
presence of solar-like oscillations.  The aim is to establish the location of 
the hot end of the stochastic convective excitation mechanism and to what
extent it may overlap the $\delta$~Scuti/$\gamma$~Doradus instability
region.  A simple but effective autocorrelation method is described which
is capable of detecting low-amplitude solar-like oscillations, but with
significant risk of a false detection.  The location of the frequency of 
maximum oscillation power, $\nu_{\rm max}$, and the large frequency separation,
$\Delta\nu$, is determined for 167 stars hotter than 6000\,K, of which 70 are 
new detections.  Results indicate that the hot edge of excitation of solar-like
oscillations does not appear to extend into the 
$\delta$~Scuti/$\gamma$~Doradus instability strip.
\end{abstract}


\section{Introduction}

Convective eddies in the outermost layers of a star have characteristic 
turn-over time scales. If the turn-over time scale matches a global 
pulsational period, energy is transferred from convective cell motion to drive
the global pulsation mode at that period; destructive interference filters out
all but the resonant frequencies. Thus random convective noise is transformed 
into distinct p-mode pulsations with a wide range of spherical harmonics.  
Stochastic oscillations driven in this way are called solar-like oscillations.  
Such oscillations were first detected in the Sun by \citet{Leighton1962} as 
quasi-periodic intensity and radial velocity variations with a period of about
5\,min.

Owing to their very small amplitudes, detection from the ground of solar-like 
oscillations in other stars had to await advances in spectroscopic detectors.  
The first unambiguous detection may perhaps be attributed to 
\citet{Kjeldsen1995b} for the G0IV star $\eta$~Boo.  The advent of space 
photometry, first with the {\it CoRoT} \citep{Fridlund2006} and later the 
{\it Kepler} spacecrafts \citep{Borucki2010}, led to the discovery of 
solar-like oscillations in thousands of stars.  Most of these are red giants 
where the oscillations have the highest amplitudes.

The individual modes in solar-like oscillations form a distinctive frequency 
pattern which is well-described by the asymptotic relation for p modes
\citep{Tassoul1980}.  Successive radial overtones, $n$, of modes with the
same spherical harmonic number, $l$, are spaced at a nearly constant
frequency interval, $\Delta\nu$, which is known as the large separation. 
The large separation depends on the mean density of the star and increases 
as the square root of the mean density.  The mode amplitudes form a distinctive 
bell-shaped envelope with maximum frequency, $\nu_{\rm max}$.  This
frequency depends on surface gravity and effective temperature and is
related to the critical acoustic frequency \citep{Brown1991}.  If 
$\nu_{\rm max}$ and $\Delta\nu$ can be measured and the effective temperature 
is known, the stellar radius and mass can be determined \citep{Stello2008,
Kallinger2010}.

The $\delta$~Scuti stars are A and early F dwarfs and giants with multiple 
frequencies higher than 5\,d$^{-1}$ while $\gamma$~Doradus stars are F 
dwarfs and giants pulsating in multiple frequencies in the range 
0.3-- 3\,d$^{-1}$.   It was originally thought that the $\delta$~Sct stars
do not pulsate with frequencies lower than about 5\,d$^{-1}$, but photometry 
from the {\it Kepler} satellite has shown that low frequencies occur in
practically all $\delta$~Sct stars \citep{Balona2018c}.  Furthermore, the 
$\gamma$~Dor stars should probably not be considered as a separate class of 
variable \citep{Xiong2016,Balona2018c}.  There are, in fact, more $\delta$~Sct stars 
than $\gamma$~Dor stars in the $\gamma$~Dor instability region.  It seems that 
the $\gamma$~Dor stars are $\delta$~Sct stars in which the high frequencies 
are damped for some unknown reason.  

\citet{Samadi2002} computed models of $\delta$~Sct stars located in the 
vicinity of the cool edge of the classical instability strip and suggested
that the amplitudes of solar-like oscillations in these stars may be
detectable even with ground-based instruments, provided they can be 
distinguished from the large-amplitude pulsations.  This is an interesting 
prospect because it would allow the stellar parameters to be determined and 
greatly simplify mode identification.  \citet{Antoci2011} reported detection 
of solar-like oscillations in the $\delta$~Sct star HD\,187547 observed by 
{\it Kepler}, but further data failed to confirm this identification (see
\citealt{Antoci2014b, Bedding2020}).  Another attempt to search for solar-like 
oscillations, this time on the $\delta$~Sct star $\rho$~Pup, also failed 
\citep{Antoci2013}.  Since then, no further searches have been published.

\citet{Huber2011b} discussed the location in the H--R diagram of stars with 
solar-like oscillations relative to the observed and theoretical cool edges of 
the $\delta$~Sct instability strip.  They do not arrive at any conclusion as 
to whether stars with solar-like oscillations are present within the 
$\delta$~Sct instability strip.

The main purpose of this investigation is to determine the location of the
hot edge of solar-like oscillations.  This is important because it places a 
constraint on any theory seeking to explain these oscillations and because it 
gives information on convection at the interface between radiative and 
convective atmospheres.  For this purpose, all stars with short-cadence 
{\it Kepler} observations in the temperature range $6000 < T_{\rm eff} < 
10000$\,K were examined for solar-like oscillations.  As a consequence, many 
new hot solar-like variables were discovered.  The hot limit of these stars
is a good estimate of the hot edge of excitation of solar-like
oscillations and places a limit on the likely amplitudes of solar-like
oscillations which might be present in $\delta$~Sct or $\gamma$~Dor stars.

\section{The data}

{\it Kepler} observations consist of almost continuous photometry of many 
thousands of stars over a four-year period with micromagnitude precision.  
The vast majority of stars were observed in long-cadence mode with a
sampling cadence of 29.4\,min. A few thousand stars were also observed in 
short-cadence mode (sampling cadence of 1\,min), but typically only for 
a few months.  The {\it Kepler} light curves used here are those with 
pre-search data conditioning (PDC) in which instrumental effects are removed 
\citep{Stumpe2012, Smith2012}.  In main sequence stars with solar-like
oscillations, the frequency of maximum amplitude, $\nu_{\rm max}$, is 
always larger than the Nyquist frequency of 24\,d$^{-1}$ of {\it Kepler} 
long-cadence mode.  Hence only short-cadence data are used.  The sample
consists of 2347 stars with $6000 < T_{\rm eff} < 10000$\,K.

Most stars in the {\it Kepler} field have  been observed by multicolour 
photometry, from which effective temperatures, surface gravities, metal 
abundances and stellar radii can be estimated. These stellar parameters are 
listed in the {\it Kepler Input Catalogue} (KIC, \citealt{Brown2011a}).  The 
effective temperatures, $T_{\rm eff}$, of {\it Kepler} stars cooler than about
6500\,K were revised by \citet{Mathur2017}.  For hotter stars, 
\citet{Balona2015d} found that adding 144\,K to the KIC $T_{\rm eff}$ 
reproduced the spectroscopic effective temperatures very well.  These 
revisions in $T_{\rm eff}$ have been used in this paper. 

\begin{figure}
\centering
\includegraphics[scale=0.18]{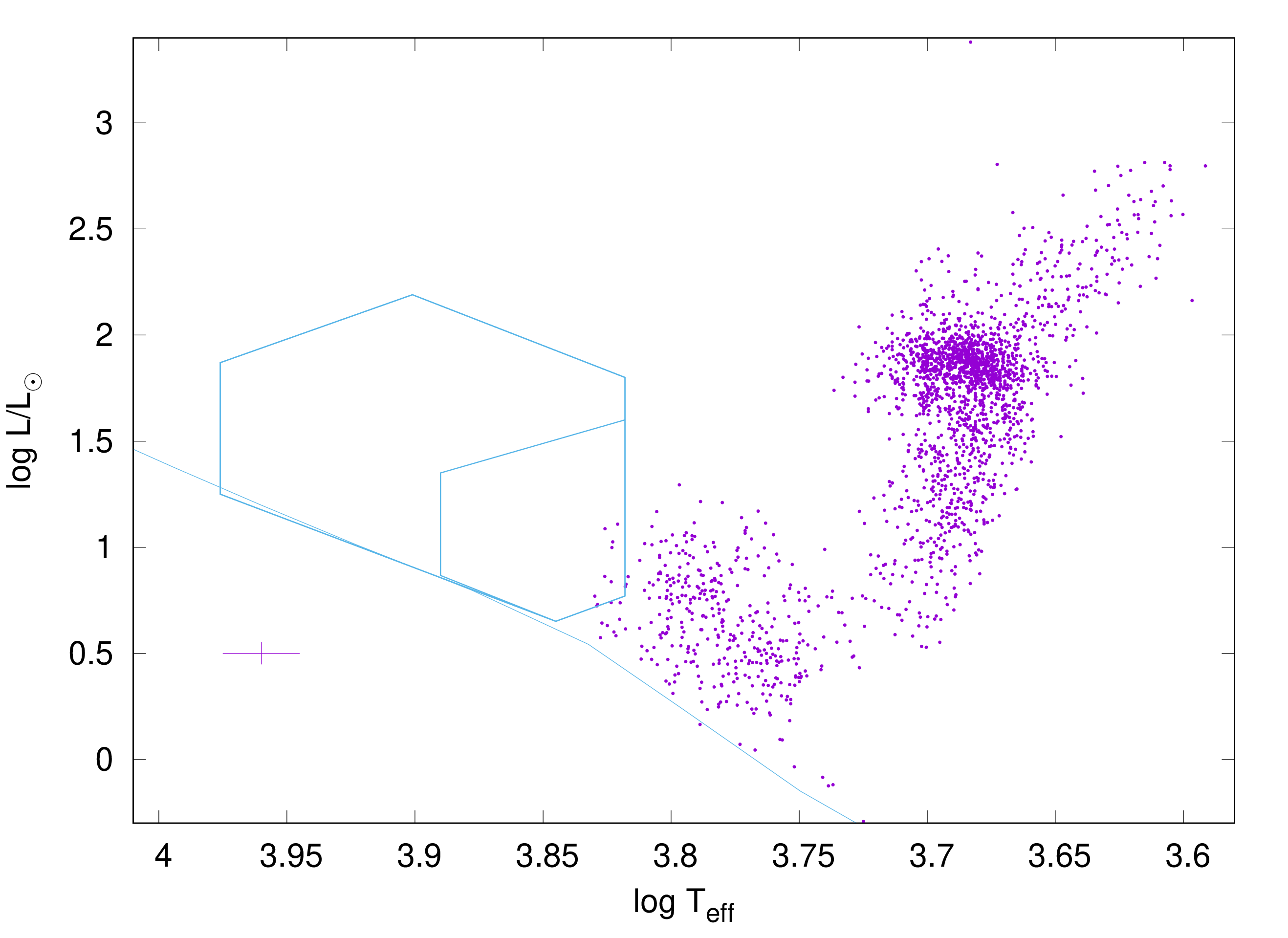}
\caption{The location of stars with solar-like oscillations from
\citet{Huber2017} in the H-R diagram.  The zero-age main sequence for models
of solar abundance from \citet{Bertelli2008} is shown (solid line).  The large 
polygonal region shows the location of $\delta$~Sct stars from 
\citet{Balona2018c}.  The smaller polygon shows the location of the 
$\gamma$~Dor stars.  The cross at bottom left indicates 1-$\sigma$ error bars in 
effective temperature and luminosity.} 
\label{huber}
\end{figure}

\citet{Huber2017} lists 2236 {\it Kepler} stars with solar-like oscillations 
compiled from various catalogues giving $\nu_{\rm max}$ and $\Delta\nu$ for
each star.  Other compilations of {\it Kepler} stars with solar-like 
oscillations are those of \citet{Chaplin2014} and \citet{Serenelli2017}.
Fig.\,\ref{huber} shows the \citet{Huber2017} stars in the Hertzprung-Russel 
(H-R) diagram together with the $\delta$~Sct and $\gamma$~Dor instability
regions from \citet{Balona2018c}.  The luminosities of these stars were 
obtained from {\em Gaia DR2} parallaxes \citep{Gaia2016, Gaia2018} using a table of 
bolometric corrections in the Sloan photometric system  by 
\citet{Castelli2003}.  Corrections for interstellar extinction were derived 
using the 3D Galactic reddening model by  \citet{Gontcharov2017}.  

It is intriguing that the hot edge of solar-like oscillations seems to  coincide 
with the cool edge of the $\delta$~Sct/$\gamma$~Dor instability regions.
Very few stars having solar-like oscillations are hotter than 6500\,K and none 
exceed 7000\,K.  

\section{Detection method}

Solar-like oscillations are usually detected by searching for the
typical bell-shaped amplitude envelope in the power spectrum of the light
curve.  Most methods use a smoothed power spectrum to obtain the global 
parameters of the Gaussian envelope.  A model consisting of the background,
which includes a contribution from granulation, and a Gaussian is then fitted 
to the data and tested for significance.  The peak of the Gaussian gives 
$\nu_{\rm max}$ while $\Delta\nu$ is estimated by autocorrelation.  This 
procedure and variants are described in, for example, \citet{Gilliland1993,
Mosser2009b, Huber2009, Mathur2010b, Benomar2012a, Lund2012}. 

All methods make use of the well-known asymptotic formula for p modes 
\citep{Tassoul1980}:
\begin{align*}
&\nu_{nl} \approx \Delta\nu\left(n + \tfrac{1}{2}l + \epsilon\right) -
l(l+1)D_0,
\end{align*}
where $l$ is the spherical harmonic number of the mode and $n$ is its radial
order.  The constant $\epsilon$ is sensitive to the surface layers and 
$D_0$ is sensitive to the sound speed gradient near the core.  This equation
is valid for non-rotating stars and leads to an approximately repetitive 
frequency pattern with characteristic frequency of $\tfrac{1}{2}\Delta\nu$.  
The factor of half comes from the fact that the $l = 1$ frequency peaks are 
approximately midway between consecutive radial ($l=0$) peaks. In a star with solar-like 
oscillations one would expect to find a  regular frequency spacing of 
$\tfrac{1}{2}\Delta\nu$  within a restricted frequency range corresponding to 
the location of bell-shaped amplitude distribution centered at 
$\nu_{\rm max}$.  Several reviews on solar-like oscillations are available
\citep{Chaplin2013, Garcia2019, Aerts2019}.

Rotation will introduce splitting of each mode into $2l + 1$ multiplets.   
However, axisymmetric modes will only be slightly affected for small to 
moderate rotation, so the repetitive frequency pattern will still
be preserved, though degraded somewhat.  Because hotter stars tend to be
more rapid rotators, the difficulty of detecting solar-like oscillations is
expected to increase with increasing effective temperature.  In addition,
the shorter mode lifetimes lead to broader frequency peaks and lower
amplitudes.

The detection of a regular frequency spacing is commonly achieved using the
power spectrum of the power spectrum of the light curve or by autocorrelation.  
The autocorrelation, $R(k)$, is the correlation of a signal with a delayed 
copy of itself with lag $k$: $R_k = \sum y(n)y(n-k)$.  The autocorrelation
can be obtained from this summation, but for computational purposes it is
faster to calculate $R_k$ from the inverse FFT of the power spectrum of the 
power spectrum of the light curve.

\begin{figure}
\centering
\includegraphics[scale=0.18]{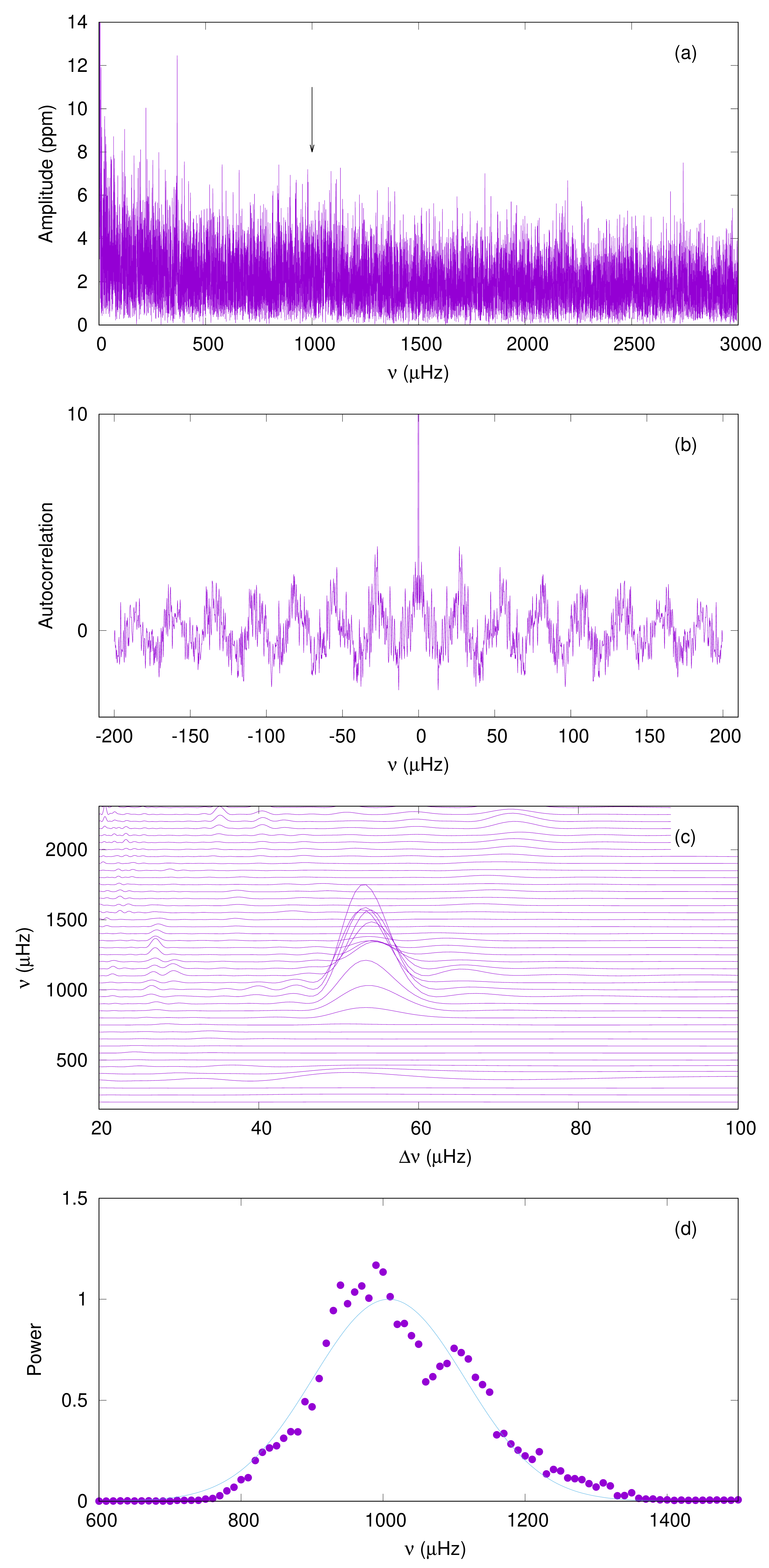}
\caption{Panel (a) shows the amplitude periodogram for KIC\,3219634, the 
arrow indicating $\nu_{\rm max}$.  Panel (b) shows the autocorrelation at a 
frequency of 1000\,$\mu$Hz showing a period of about $\tfrac{1}{2}\Delta\nu 
\approx 27$\,$\mu$Hz.  Panel (c) shows power spectra of the autocorrelation 
(after removing the central peak) plotted against frequency  (the 
$\nu$ -- $\Delta\nu$ diagram). Panel (d) shows the maximum autocorrelation 
power as a function of frequency within a 10 percent range of the approximate
$\Delta\nu$. The curve is a fitted Gaussian with a maximum at 
$\nu_{\rm max} = 1008.1 \pm 2.8$\,$\mu$Hz.  The average $\Delta\nu$ within
$\pm\tfrac{1}{2}$FWHM of the maximum is $54.06 \pm 0.18$\,$\mu$Hz.}
\label{solar}
\end{figure}

The method used here most closely resembles that of \citet{Mathur2010b} in 
that the presence of equally-spaced frequencies in some region of the 
power spectrum of the light curve is detected.  A suitably small region of
the power spectrum of the light curve is selected and the autocorrelation
function is calculated.  If a pattern of equally-spaced frequencies exists
in this region, the autocorrelation function will display a periodic
variation with period equal to the frequency separation.  The amplitude of
the periodic variation depends on the strength of the correlation and follows 
the bell shape seen in the amplitude envelope of the solar-like oscillations
in the power spectrum of the light curve.

The frequency range, $\delta\nu$, of the portion of the power spectrum of the 
light curve used to calculate the autocorrelation needs careful consideration.
We know that $\Delta\nu$ is roughly proportional to $\nu_{\rm max}$ (more
accurately, $\Delta\nu \propto \nu_{\rm max}^s$ where $s \approx 0.8$,
\citealt{Stello2009b}).  At high frequencies $\delta\nu$ needs to be large 
enough to include a few periods in the autocorrelation function.  Thus
$\delta\nu$ should be at least 2 or 3 times $\Delta\nu$.  At low frequencies
$\Delta\nu$ will be smaller and hence $\delta\nu$ can be made correspondingly 
smaller. In other words, the frequency range to be sampled for autocorrelation 
should be roughly proportional to the frequency, $\nu$, that is being
sampled, i.e. $\delta\nu = \alpha\nu$.  A value of $\alpha =$ 0.2--0.4 was 
found to be suitable.

To measure the period of the autocorrelation function, it is convenient to
calculate the power spectrum of the autocorrelation.  This is actually the
same as the power spectrum of the power spectrum of the light curve. 
However, it was found that removing the large maximum at zero lag, which
is always present in the autocorrelation, leads to a cleaner power spectrum.  
The period of variation in the autocorrelation, if any exists, is given by the 
inverse of the frequency of the highest peak in the power spectrum of the
modified autocorrelation.  Its amplitude measures the significance of the
period.  

In the method used here, the autocorrelation is calculated at equal
frequency steps in the power spectrum of the light curve.  Let $\nu$ be the 
central frequency at which the autocorrelation is calculated.  The power
spectrum of the autocorrelation function is plotted at this frequency. If a
solar-like oscillation is present, a significant peak will occur in this
power spectrum at a period given by $\tfrac{1}{2}\Delta\nu$.  By plotting
the power spectra at each sampled frequency, $\nu$, a solar oscillation that
may be present will be revealed because the maxima of the power spectra will
occur at approximately the same period in the power spectrum.  It is
convenient to plot $\Delta\nu$ itself instead of the period.  This
$\nu$ -- $\Delta\nu$ plot is central to detection of solar-like
oscillations.

The value of $\nu_{\rm max}$ will be the value of $\nu$ at which the peak 
attains maximum amplitude.  This can be obtained by sampling the power
spectrum of the light curve with smaller frequency steps on a second pass.
At each frequency step, the maximum power of the autocorrelation is plotted 
as a function of $\nu$.  This relationship is very close to a Gaussian in 
most cases, so $\nu_{\rm max}$ may be obtained by fitting a Gaussian to this 
curve.  The uncertainty in $\nu_{\rm max}$ is taken to be the standard deviation of 
the location of the Gaussian peak obtained from the least-squares fit.
Since $\Delta\nu$ is only approximately constant, a best estimate of
$\Delta\nu$ is found by taking the mean value within the full-width at half
maximum of the Gaussian centered on $\nu_{\rm max}$.  The uncertainty in 
$\Delta\nu$ is taken as the standard deviation of these values.

After a suitable list of candidate stars is compiled and the Lomb-Scargle
power spectra calculated, the power spectrum of the modified correlation is
plotted as a function of $\Delta\nu$ (the $\nu$ -- $\Delta\nu$ plot) using a 
relatively large frequency step in $\nu$.  This plot is examined for the 
possible presence of a significant peak.  As an example, Fig.\,\ref{solar}(a) 
shows the amplitude periodogram of the light curve of KIC\,3219634, which is 
a previously undiscovered solar-like pulsator. The solar-like oscillations are
hardly detectable by visual inspection of the periodogram. Fig.\,\ref{solar}(b) 
shows the autocorrelation function near $\nu_{\rm max}$.  The periodicity is 
clearly seen.  Fig\,\ref{solar}(c) shows part of the $\nu$ - $\Delta\nu$ 
diagram calculated with a step size of 50\,$\mu$Hz in $\nu$.  It is clear 
that equally-spaced frequencies are present at about $\Delta\nu \approx 
55$\,$\mu$Hz and $\nu_{\rm max} \approx$ 1000\,$\mu$Hz.   In a second pass,
the step size in $\nu$ has been reduced to 10\,$\mu$Hz and at each frequency
step the maximum power in the modified autocorrelation is measured within a
restricted range of $\Delta\nu$.   The range was taken to be
$\pm0.1\Delta\nu$.  A Gaussian fitted to the larger peak gives the best
estimate of $\nu_{\rm max}$.

The detection of the ``humps'' which represent a region of high
autocorrelation is never a problem as they always have high signal-to-noise. 
The example of Fig.\,\ref{solar}(c) is fairly typical.  The main problem is
that for some stars many similar humps and/or ridges are visible.  For
example, eclipsing binaries have significant power at large harmonics which
appear as ridges in the $\nu$ --- $\Delta\nu$ diagram.  Although these cannot 
be mistaken for the hump characteristic of solar-like oscillations, it 
interferes with detection of such a hump. Pulsating stars, such as 
$\delta$~Sct/$\gamma$~Dor variables also have ridges in the diagram which tend 
to confuse the detection of solar-like oscillations.  This is the case for 
HD\,187547, the $\delta$~Sct star previously thought to contain solar-like 
oscillations \citep{Antoci2011}.  This means that any solar-like oscillations 
in $\delta$~Sct stars, if they exist, may be obscured by the ridges created
by $\delta$~Sct pulsation in the $\nu$ -- $\Delta\nu$ diagram.  If a hump lies 
close to the expected location in ($\nu_{\rm max}, \Delta\nu$), it may 
represent a true oscillation, but one cannot be certain.  The method is
aimed at finding oscillations with very low amplitudes.  This, of course, is 
what is needed to extend the region where these oscillations are to be found 
to higher effective temperatures.  The price to be paid is that there is a 
significant risk of a false detection.

\section{Results}

Of the 2347 stars with $6000 < T_{\rm eff} < 10000$\,K observed in 
short-cadence mode, solar-like oscillations were detected in 167 stars.  
This includes the 97 stars in Table 1 (in the Appendix) which were already 
known to have solar-like oscillations.  The fact that all stars with 
$T_{\rm eff} > 6000$\,K and with known solar-like oscillations were quite 
easily detected, lends confidence to the method described above.  In addition, 
Table\,2 in the Appendix lists 70 stars with newly-found solar-like 
oscillations.  For all 167 stars, there is no difficulty in locating the hump 
as it is the only one with significant visibility.  The tables give the
oscillation parameters as well as the adopted effective temperatures and
luminosities derived from {\em Gaia DR2} parallaxes as mentioned above.  The 
median rms error is 1.95\,$\mu$Hz in $\nu_{\rm max}$ and 0.31\,$\mu$Hz in 
$\Delta\nu$. 

\begin{figure}
\centering
\includegraphics[scale=0.18]{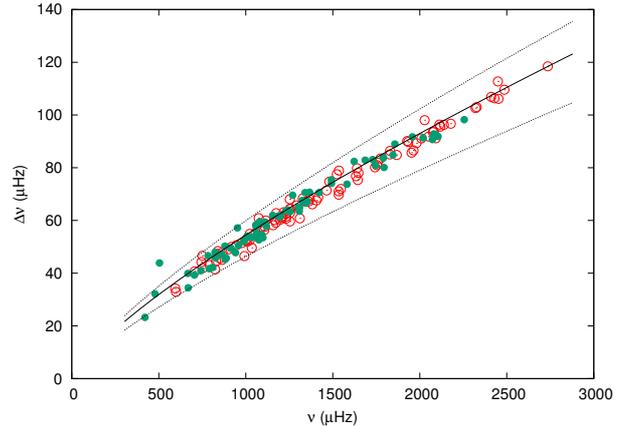}
\caption{The large separation, $\Delta\nu$, as a function of $\nu_{\rm max}$ 
for 167 stars.  Stars known from the literature are shown by open circles
(red), newly detected stars by filled circles (green).  The solid line is the
relation from \citet{Stello2009b} and dotted lines show +10 and -15 per cent 
deviations.}
\label{nudnu}
\end{figure}

\begin{figure}
\centering
\includegraphics[scale=0.18]{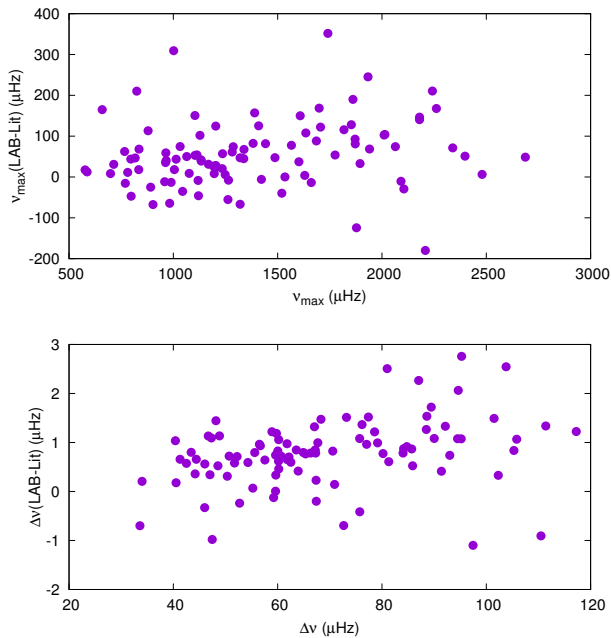}
\caption{Comparison of $\nu_{\rm max}$ (top panel) and $\Delta\nu$
(bottom panel) obtained in this paper (LAB) with those in the literature.}
\label{hublab}
\end{figure}

Fig.\,\ref{nudnu} shows the large separation, $\Delta\nu$, as a 
function of $\nu_{\rm max}$ for the 167 stars.  The 70 newly discovered
stars follow the same relationship as the 97 stars in which solar-like 
oscillations are known to be present.  This is a strong indication that the 
detected oscillations are real.  

Fig.\,\ref{hublab} compares the values of $\nu_{\rm max}$ and $\Delta\nu$ 
obtained here with those in the literature.  In both cases, the difference
increases with $\nu_{\rm max}$.  Neither $\nu_{\rm max}$ or $\Delta\nu$ are 
well-defined quantities.  $\nu_{\rm max}$ is usually estimated by fitting a 
Gaussian to the envelope of the peaks.  $\Delta\nu$ varies slightly with 
$\nu$ and is normally measured at $\nu_{\rm max}$.  In both cases the trend
shown in Fig.\,\ref{hublab} may be related to the fact that the frequency
range, $\delta\nu$, which is sampled to derive the autocorrelation increases
with $\nu_{\rm max}$, as described above.  In short, the method described here 
is suited to the detection of low-amplitude signals but is not optimized for a 
strict evaluation of $\nu_{\rm max}$ and $\Delta\nu$.  This clearly requires 
individual peaks to be resolved. 

\begin{figure}
\centering
\includegraphics[scale=0.18]{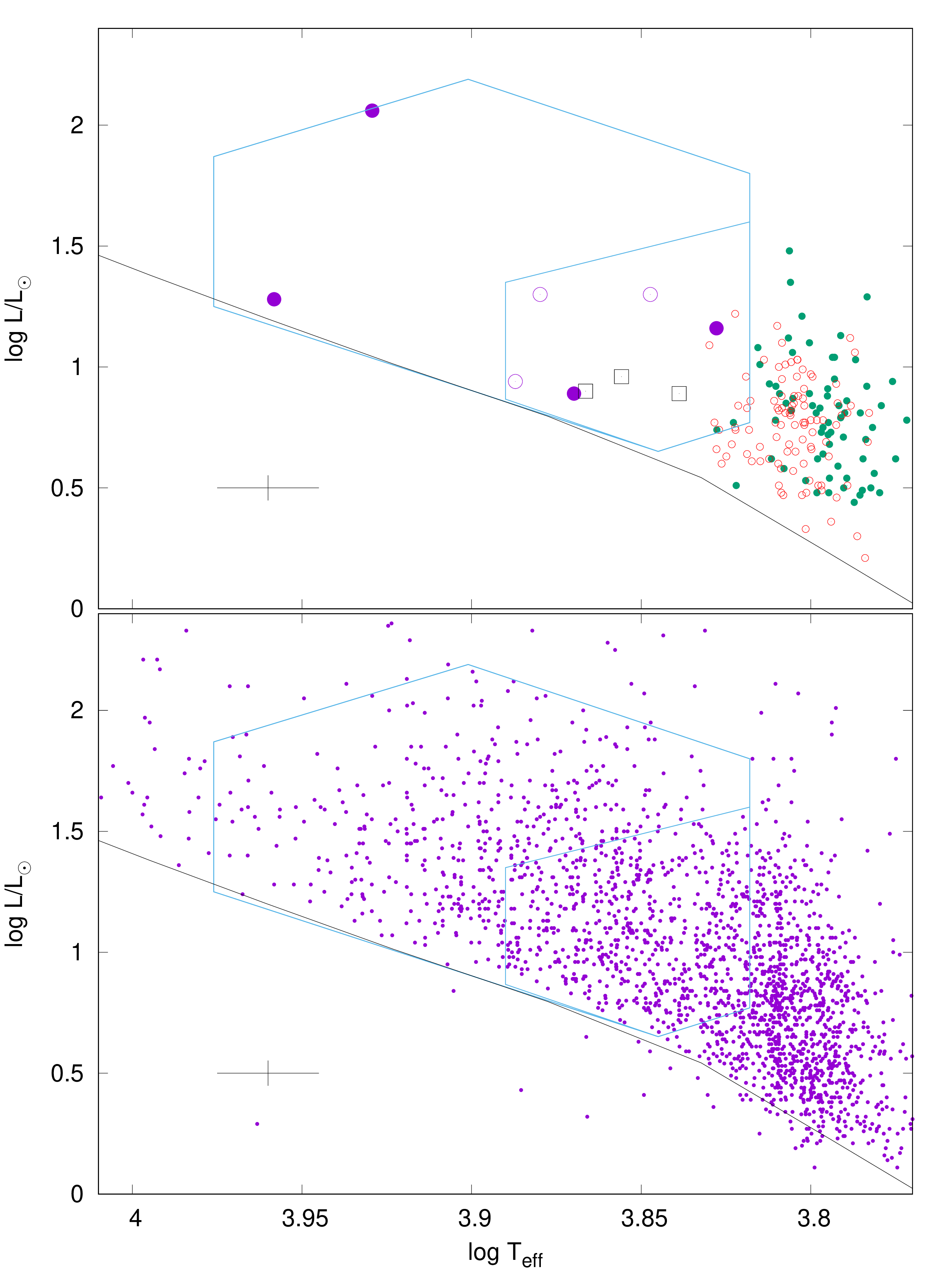}
\caption{Top panel: the location in the H--R diagram of stars in Tables
1--3. Newly detected stars are shown by the small filled 
green circles and known stars by small open red circles.  The large symbols
are stars with possible solar-like oscillations: violet filled circles - 
non-variable stars; violet open circles: $\delta$~Sct; black squares - 
$\gamma$~Dor stars.   Bottom panel: the small filled circles are from the 
sample of 2347 stars used to search for solar-like oscillations.  The zero-age 
main sequence for models of solar abundance from \citet{Bertelli2008} is shown 
(black solid line).  The large polygonal region shows the location of 
$\delta$~Sct stars from \citet{Balona2018c}.   The  smaller polygon shows the 
location of the $\gamma$~Dor stars.  The cross at bottom left shows 
1-$\sigma$ error bars.} 
\label{hrsol}
\end{figure}

\begin{figure}
\centering
\includegraphics[scale=0.18]{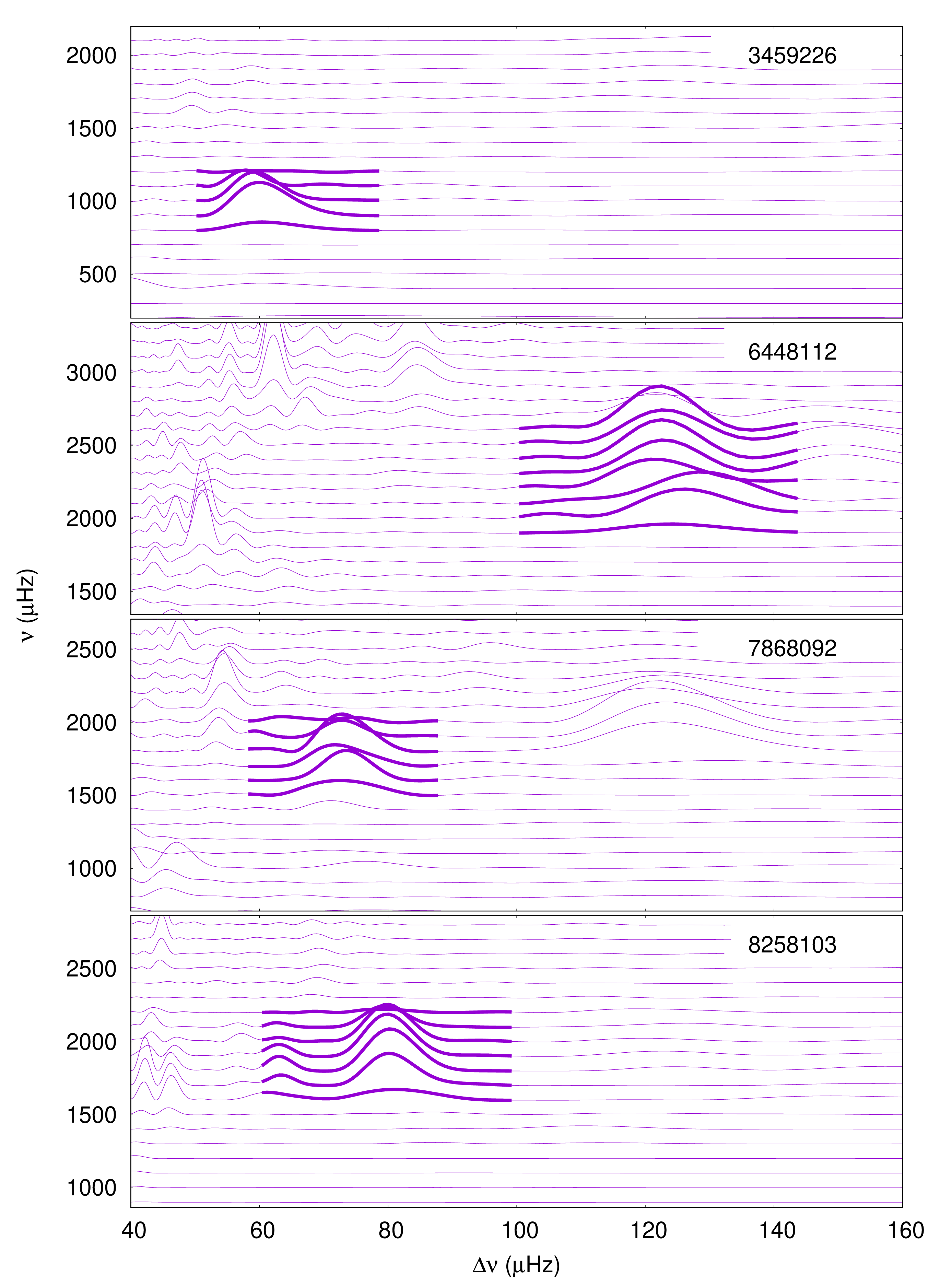}
\caption{The $\nu$ -- $\Delta\nu$ diagrams for the four hottest
non-pulsating stars. The thicker solid lines indicate the region which might
be a result of solar-like oscillations.}
\label{nonvar}
\end{figure}

The top panel in Fig.\,\ref{hrsol} shows the location of the 167 stars 
with solar-like oscillations in the H-R diagram as well as the $\delta$~Sct
and $\gamma$~Dor region of instability from \citet{Balona2018c}.  The bottom
panel shows the location of the 2347 stars in our sample.  This is to show
that there is a uniform decrease of density of stars with increasing $T_{\rm
eff}$.  There is no abrupt change in stellar density which might be
responsible for a sharp decrease in number of stars with solar-like
oscillations.  In the effective temperature range 6000---6500\,K about
17\,percent of stars in our sample are solar-like oscillators.  If this
fraction remains the same for hotter stars, at least 80 solar-like
oscillators may be expected just within the $\delta$~Sct instability strip
in the temperature range 6500--7000\,K.

Two apparently non-pulsating stars KIC\,6448112 ($T_{\rm eff} = 8498$\,K) and 
KIC\,8258103 ($T_{\rm eff} = 9082$\,K) seem to have a hump in the 
$\nu$ -- $\Delta\nu$ diagram which might be a result of solar-like
oscillations.  KIC\,7868092 ($T_{\rm eff} = 6726$\,K) and KIC\,3459226 
($T_{\rm eff} = 7410$\,K) are somewhat cooler non-pulsating stars in which a 
hump is also visible.  In all these stars, no indication of periodogram peaks 
are seen at $\nu_{\rm max}$.

The $\nu$ -- $\Delta\nu$ diagrams for these stars are shown in 
Fig.\,\ref{nonvar}.  In every case the detection of possible solar-like 
oscillations is very tentative because of the presence of similar hump-like 
features.  For this reason, it is not certain that solar-like oscillations are 
present, even though the measured $\nu_{\rm max}$ and $\Delta\nu$ are
consistent with solar-like oscillations.

Indications of solar like oscillations were also detected in three 
$\delta$~Sct stars (KIC\,7212040, KIC\.5476495, KIC\,9347095), and three 
$\gamma$~Dor stars (KIC\,9696853, KIC\,8264254, KIC\,5608334).  Once again, 
many other hump-like features are present as well as ridges in the 
$\nu$ -- $\Delta\nu$ diagram.  As before, this greatly reduces the
confidence that solar-like oscillations are present.  For the sake of 
completeness, oscillation parameters of these ten stars are given in
Table\,3 in the Appendix.

\section{Discussion and conclusions}

In this study an attempt is made to detect solar-like oscillating stars
hotter than 6000\,K in order to determine the hot edge of the excitation
of solar-like oscillations in main sequence stars.  The temperature 
region examined includes the $\delta$~Sct/$\gamma$~Dor instability strip.
Discovery of solar-like oscillations in $\delta$~Sct stars would, of course,
be of great interest.  There are also large numbers of non-pulsating stars
within the $\delta$~Sct/$\gamma$~Dor instability strip.  Detecting
solar-like oscillations in these stars would be much easier than in
$\delta$~Sct or $\gamma$~Dor stars because the low-amplitude of solar-like
oscillations is likely to be swamped by the much higher amplitudes of
self-driven modes.
 
In order to improve the detection probability, a simple method of
detecting solar-like oscillations of low amplitudes (well below the level of
detection by inspection of the periodogram) was devised.  However, the
method does have a significant risk of a false detection. The method does
not require smoothing of the periodogram or fitting of a granulation model.
In this method, a search is made for equal frequency spacings using 
autocorrelation.  It is found that the power spectrum of the autocorrelation 
is a powerful tool which easily allows detection of very low-amplitude 
solar-like oscillations. This is accomplished by inspecting a plot of the 
power spectrum as a function of frequency (the $\nu$ -- $\Delta\nu$ diagram).  
In this way, all known solar-like oscillating stars hotter than 6000\,K were 
independently detected without difficulty.

Application of this method to stars with $6000 < T_{\rm eff} < 10000$\,K led 
to the detection of 70 previously unknown solar-like oscillating stars. None 
of these stars are located within the $\delta$~Sct/$\gamma$~Dor instability 
strip.  If the fraction of stars with solar-like oscillations is the same on
either side of the cool edge of the $\delta$~Sct instability strip, then
several dozen such stars should have been detected.  There are 77 stars within 
the $\delta$~Sct instability strip with no apparent variability.  These would 
be the best candidates as there is no interference from self-driven
pulsations.  Yet solar-like oscillations are not found with certainty in any
of these stars.

Some uncertain indication of solar like oscillations were detected in four
non-variable stars and six pulsating stars inside the $\delta$~Sct
instability strip.  However, the detections are not sufficiently convincing.
None of these ten stars constitute sufficient proof that the hot edge of 
solar-like oscillations extends to within the $\delta$~Sct/$\gamma$~Dor 
instability strip.

It should be noted that the line-widths in the periodogram of main sequence
stars with solar-like oscillations increases with effective temperature
\citep{White2012, Lund2017, Compton2019}.  This is a consequence of the
shorter stochastic mode lifetimes.  Increase of line broadening results in a
decrease of line amplitude for oscillations of the same energy.  In
addition, the rotation rate increases with $T_{\rm eff}$, further
complicating the frequency spectrum.  As a result, detection of individual 
modes in these stars becomes increasingly more difficult with increasing
$T_{\rm eff}$.  However, the characteristic bell-shaped envelope of
solar-like oscillations should still be visible.  The author has inspected
the periodograms of many thousands of $\delta$~Sct stars, but no indication
of such a feature has ever been seen.

The conclusion is that the hot edge of solar-like oscillations probably
coincides with the cool edge of the $\delta$~Sct/$\gamma$~Dor instability 
strip.  This is quite an interesting coincidence and raises the question of 
whether the physical factors giving rise to self-driven pulsations also act to 
damp solar-like oscillations.  Further study of this problem would require the 
precision and time span matching that of the {\it Kepler} photometry.  
Hopefully, this may be attained in the not too distant future.

\section*{Acknowledgments}

LAB wishes to thank the National Research Foundation of South Africa for 
financial support. This paper includes data collected by the Kepler 
mission. Funding for the Kepler mission is provided by the NASA 
Science Mission directorate.  This work also used data from the European 
Space Agency (ESA) mission Gaia (\url{https://www.cosmos.esa.int/gaia}), 
processed by the Gaia Data Processing and Analysis Consortium (DPAC,
\url{https://www.cosmos.esa.int/web/gaia/dpac/consortium}). Funding for 
the DPAC has been provided by national institutions, in particular the 
institutions participating in the Gaia Multilateral Agreement.

\bibliographystyle{mnras}
\bibliography{gsolar}

\newpage

\section*{Appendix}

\begin{table*}
\begin{center}
\caption{Table of known stars with solar-like oscillations.  The first column is the 
KIC number.  The values of $\nu_{\rm max}$ ($\mu$Hz) and $\Delta\nu$ ($\mu$Hz) 
(together with their 1-\,$\sigma$ errors) follows.  The effective temperature,
$T_{\rm eff}$ (K), from the literature and the luminosity, $\log{L/L_\odot}$,
from the {\em Gaia DR2} parallax, are shown.  Note that for stars with
$T_{\rm eff} > 6500$\,K, values are from the calibration of
\citet{Balona2015d}.}

\vspace{3mm}

\tiny

\begin{tabular}{rrrrrrrrrrrrr}
\hline
\multicolumn{1}{c}{KIC}                  & 
\multicolumn{1}{c}{$\nu_{\rm max}$}      & 
\multicolumn{1}{c}{$\Delta\nu$}          &
\multicolumn{1}{c}{$T_{\rm eff}$}         & 
\multicolumn{1}{c}{$\log\tfrac{L}{L_\odot}$}  & 
\multicolumn{1}{c}{KIC}                  & 
\multicolumn{1}{c}{$\nu_{\rm max}$}      & 
\multicolumn{1}{c}{$\Delta\nu$}          &
\multicolumn{1}{c}{$T_{\rm eff}$}         & 
\multicolumn{1}{c}{$\log\tfrac{L}{L_\odot}$}  \\
\hline
   1430163 & $ 1829.0 \pm  3.6$ & $  86.39 \pm  0.21$ &  6588 &   0.64 &  8367710 & $ 1107.1 \pm  1.2$ & $  56.40 \pm  0.25$ &  6364 &   0.88 \\
   1435467 & $ 1464.6 \pm  1.5$ & $  71.38 \pm  0.40$ &  6306 &   0.66 &  8377423 & $ 1023.0 \pm  2.3$ & $  52.46 \pm  0.23$ &  6141 &   0.84 \\
   1725815 & $ 1009.0 \pm  1.7$ & $  51.96 \pm  0.15$ &  6196 &   0.85 &  8408931 & $  594.3 \pm  1.8$ & $  34.20 \pm  0.69$ &  6143 &   1.12 \\
   2837475 & $ 1633.5 \pm  1.1$ & $  76.80 \pm  0.12$ &  6641 &   0.74 &  8420801 & $ 1361.1 \pm  3.1$ & $  67.20 \pm  0.38$ &  6245 &   0.69 \\
   3123191 & $ 2092.3 \pm  6.7$ & $  91.16 \pm  0.60$ &  6450 &   0.60 &  8494142 & $ 1174.9 \pm  0.7$ & $  62.77 \pm  0.23$ &  6070 &   0.69 \\
   3236382 & $ 1643.5 \pm  3.9$ & $  75.33 \pm  0.65$ &  6715 &   0.74 &  8579578 & $  946.8 \pm  1.2$ & $  50.64 \pm  0.08$ &  6339 &   0.91 \\
   3344897 & $  835.0 \pm  1.8$ & $  47.35 \pm  0.12$ &  6434 &   0.95 &  8866102 & $ 2118.0 \pm  1.4$ & $  95.57 \pm  0.30$ &  6264 &   0.51 \\
   3424541 & $  754.0 \pm  1.0$ & $  41.94 \pm  0.07$ &  6123 &   1.06 &  8940939 & $ 1535.1 \pm  3.3$ & $  78.91 \pm  0.25$ &  6342 &   0.77 \\
   3456181 & $ 1005.9 \pm  1.1$ & $  52.93 \pm  0.29$ &  6381 &   0.88 &  8956017 & $ 1254.5 \pm  4.1$ & $  62.87 \pm  0.58$ &  6397 &   0.80 \\
   3547794 & $ 1327.7 \pm  4.5$ & $  68.37 \pm  0.54$ &  6462 &   0.77 &  9116461 & $ 2452.2 \pm  3.7$ & $ 106.13 \pm  0.39$ &  6429 &   0.47 \\
   3633889 & $ 2113.0 \pm  4.3$ & $  96.35 \pm  0.60$ &  6328 &   0.48 &  9139151 & $ 2735.7 \pm  2.6$ & $ 118.47 \pm  0.17$ &  6114 &   0.30 \\
   3733735 & $ 2075.7 \pm  1.9$ & $  93.47 \pm  0.13$ &  6726 &   0.66 &  9139163 & $ 1757.8 \pm  1.3$ & $  81.90 \pm  0.40$ &  6403 &   0.68 \\
   3967430 & $ 1981.1 \pm  2.9$ & $  89.30 \pm  0.39$ &  6682 &   0.63 &  9163769 & $ 1753.3 \pm  4.1$ & $  80.97 \pm  1.47$ &  6410 &   0.65 \\
   4465529 & $ 1546.7 \pm  3.0$ & $  71.96 \pm  0.87$ &  6344 &   0.70 &  9206432 & $ 1953.0 \pm  1.6$ & $  85.69 \pm  0.21$ &  6531 &   0.67 \\
   4586099 & $ 1229.6 \pm  1.3$ & $  62.47 \pm  0.46$ &  6378 &   0.76 &  9226926 & $ 1480.3 \pm  1.6$ & $  74.71 \pm  0.57$ &  6736 &   0.77 \\
   4638884 & $ 1231.3 \pm  1.1$ & $  61.44 \pm  0.14$ &  6448 &   0.82 &  9287845 & $ 1034.7 \pm  3.5$ & $  49.60 \pm  0.43$ &  6309 &   0.97 \\
   4646780 & $ 1074.5 \pm  3.2$ & $  60.85 \pm  0.57$ &  6628 &   0.84 &  9328372 & $ 1406.0 \pm  5.4$ & $  67.59 \pm  0.56$ &  6390 &   0.84 \\
   4931390 & $ 2080.4 \pm  1.2$ & $  93.73 \pm  0.12$ &  6703 &   0.60 &  9455860 & $ 1534.3 \pm  3.4$ & $  71.05 \pm  0.93$ &  6579 &   0.74 \\
   5095850 & $  750.6 \pm  7.5$ & $  46.59 \pm  0.59$ &  6759 &   1.09 &  9457728 & $ 1114.4 \pm  1.9$ & $  59.97 \pm  0.49$ &  6284 &   0.78 \\
   5214711 & $ 1310.9 \pm  1.7$ & $  60.78 \pm  0.43$ &  6302 &   0.73 &  9542776 & $  823.8 \pm  3.6$ & $  41.42 \pm  0.51$ &  6435 &   1.10 \\
   5431016 & $  917.8 \pm  0.9$ & $  49.97 \pm  0.26$ &  6594 &   0.96 &  9697131 & $ 1204.2 \pm  2.0$ & $  61.25 \pm  0.29$ &  6337 &   0.84 \\
   5516982 & $ 1827.7 \pm  1.4$ & $  85.02 \pm  0.47$ &  6386 &   0.57 &  9812850 & $ 1292.9 \pm  1.3$ & $  65.70 \pm  0.39$ &  6458 &   0.71 \\
   5636956 & $ 1056.8 \pm  2.0$ & $  55.30 \pm  0.14$ &  6436 &   0.88 &  9821513 & $ 1230.8 \pm  1.7$ & $  60.66 \pm  0.12$ &  6335 &   0.77 \\
   5773345 & $ 1156.7 \pm  1.0$ & $  58.19 \pm  0.28$ &  6179 &   0.80 & 10003270 & $  792.3 \pm  2.5$ & $  43.07 \pm  0.15$ &  6395 &   1.02 \\
   5961597 & $ 1206.5 \pm  1.9$ & $  60.94 \pm  0.11$ &  6573 &   0.86 & 10016239 & $ 2326.8 \pm  1.8$ & $ 102.98 \pm  0.40$ &  6277 &   0.51 \\
   6064910 & $  828.9 \pm  1.5$ & $  44.53 \pm  0.23$ &  6370 &   0.96 & 10024648 & $ 1084.7 \pm  1.6$ & $  57.46 \pm  0.39$ &  6258 &   0.78 \\
   6225718 & $ 2410.3 \pm  1.4$ & $ 106.89 \pm  0.12$ &  6223 &   0.36 & 10070754 & $ 2320.4 \pm  3.1$ & $ 102.62 \pm  0.38$ &  6437 &   0.48 \\
   6232600 & $ 2178.2 \pm  2.0$ & $  96.68 \pm  0.32$ &  6447 &   0.51 & 10081026 & $  745.0 \pm  2.8$ & $  44.21 \pm  0.08$ &  6368 &   1.03 \\
   6508366 & $  998.0 \pm  0.9$ & $  52.31 \pm  0.24$ &  6343 &   0.87 & 10273246 & $  903.6 \pm  1.2$ & $  49.07 \pm  0.38$ &  6155 &   0.81 \\
   6530901 & $ 1639.1 \pm  3.2$ & $  79.79 \pm  0.71$ &  6233 &   0.63 & 10351059 & $ 1382.1 \pm  3.4$ & $  66.13 \pm  0.27$ &  6435 &   0.83 \\
   6612225 & $ 1167.1 \pm  1.4$ & $  60.12 \pm  0.12$ &  6309 &   0.78 & 10355856 & $ 1366.9 \pm  1.7$ & $  68.06 \pm  0.22$ &  6438 &   0.76 \\
   6679371 & $  976.1 \pm  0.8$ & $  51.42 \pm  0.07$ &  6302 &   0.96 & 10454113 & $ 2428.7 \pm  1.3$ & $ 106.34 \pm  0.24$ &  6155 &   0.51 \\
   7103006 & $ 1199.5 \pm  1.0$ & $  60.34 \pm  0.26$ &  6420 &   0.81 & 10491771 & $ 1773.9 \pm  6.6$ & $  83.50 \pm  0.96$ &  6491 &   0.62 \\
   7106245 & $ 2448.9 \pm  2.3$ & $ 112.73 \pm  0.21$ &  6082 &   0.21 & 10709834 & $ 1416.2 \pm  1.5$ & $  68.70 \pm  0.11$ &  6453 &   0.83 \\
   7133688 & $ 1172.7 \pm  3.5$ & $  59.09 \pm  0.23$ &  6383 &   0.85 & 10730618 & $ 1343.2 \pm  1.1$ & $  67.11 \pm  0.34$ &  6401 &   0.81 \\
   7206837 & $ 1742.7 \pm  1.3$ & $  80.14 \pm  0.42$ &  6340 &   0.60 & 10972252 & $  863.4 \pm  1.9$ & $  45.05 \pm  0.27$ &  6202 &   0.93 \\
   7218053 & $  707.0 \pm  1.9$ & $  40.70 \pm  0.46$ &  6366 &   1.03 & 11070918 & $ 1253.2 \pm  1.5$ & $  68.04 \pm  0.49$ &  6642 &   1.22 \\
   7282890 & $  852.0 \pm  1.4$ & $  45.67 \pm  0.12$ &  6345 &   0.99 & 11081729 & $ 2009.0 \pm  1.8$ & $  91.10 \pm  0.50$ &  6530 &   0.61 \\
   7465072 & $ 1534.2 \pm  3.9$ & $  69.77 \pm  1.35$ &  6309 &   0.67 & 11189107 & $  992.2 \pm  4.8$ & $  46.47 \pm  0.46$ &  6513 &   1.03 \\
   7529180 & $ 1963.7 \pm  2.0$ & $  86.56 \pm  0.20$ &  6657 &   0.68 & 11229052 & $ 1867.5 \pm  2.9$ & $  84.82 \pm  0.17$ &  6376 &   0.65 \\
   7530690 & $ 2027.9 \pm  4.4$ & $  98.02 \pm  0.57$ &  6568 &   0.61 & 11253226 & $ 1647.5 \pm  1.2$ & $  78.01 \pm  0.15$ &  6642 &   0.75 \\
   7622208 & $ 1254.0 \pm  6.9$ & $  59.60 \pm  0.69$ &  6396 &   0.82 & 11453915 & $ 2486.2 \pm  4.5$ & $ 109.58 \pm  1.01$ &  6331 &   0.33 \\
   7670943 & $ 1927.8 \pm  1.4$ & $  90.13 \pm  0.20$ &  6315 &   0.53 & 11460626 & $  839.8 \pm  1.9$ & $  48.39 \pm  0.28$ &  6065 &   0.81 \\
   7800289 & $  599.0 \pm  3.8$ & $  32.89 \pm  0.42$ &  6455 &   1.17 & 11467550 & $ 1023.0 \pm  2.9$ & $  54.89 \pm  0.35$ &  6202 &   0.76 \\
   7938112 & $ 1218.4 \pm  4.5$ & $  63.12 \pm  1.03$ &  6337 &   0.76 & 11757831 & $ 1256.3 \pm  2.7$ & $  64.35 \pm  0.43$ &  6468 &   0.86 \\
   8150065 & $ 1933.7 \pm  1.6$ & $  89.76 \pm  0.17$ &  6263 &   0.49 & 11919192 & $  866.2 \pm  1.8$ & $  47.82 \pm  0.33$ &  6448 &   1.00 \\
   8179536 & $ 2138.5 \pm  1.6$ & $  96.27 \pm  0.28$ &  6347 &   0.47 & 12156916 & $ 1523.4 \pm  1.9$ & $  77.53 \pm  0.24$ &  6438 &   0.58 \\
   8216936 & $ 1110.1 \pm  2.6$ & $  57.61 \pm  0.38$ &  6419 &   1.01 & 12317678 & $ 1254.3 \pm  1.3$ & $  64.42 \pm  0.11$ &  6589 &   0.83 \\
   8298626 & $ 2050.9 \pm  3.3$ & $  91.81 \pm  0.20$ &  6200 &   0.46 &          &                    &                     &       &        \\
\hline
\end{tabular}
\end{center}
\end{table*}

\begin{table*}
\begin{center}
\caption{Newly-discovered stars with solar-like oscillations.  The cloumns
are the same as Table~1.}

\vspace{3mm}

\tiny
\begin{tabular}{rrrrrrrrrrrrr}
\hline
\multicolumn{1}{c}{KIC}                  & 
\multicolumn{1}{c}{$\nu_{\rm max}$}      & 
\multicolumn{1}{c}{$\Delta\nu$}          &
\multicolumn{1}{c}{$T_{\rm eff}$}         & 
\multicolumn{1}{c}{$\log\tfrac{L}{L_\odot}$}  & 
\multicolumn{1}{c}{KIC}                  & 
\multicolumn{1}{c}{$\nu_{\rm max}$}      & 
\multicolumn{1}{c}{$\Delta\nu$}          &
\multicolumn{1}{c}{$T_{\rm eff}$}         & 
\multicolumn{1}{c}{$\log\tfrac{L}{L_\odot}$}  \\
\hline
   3102595 & $ 1074.9 \pm  3.3$ & $  52.65 \pm  0.18$ &  6183 &   0.79 &   8279146 & $ 1366.7 \pm  3.2$ & $  70.69 \pm  0.34$ &  6650 &   0.77 \\
   3219634 & $ 1008.1 \pm  2.7$ & $  54.06 \pm  0.18$ &  6286 &   0.81 &   8346342 & $  915.8 \pm  0.8$ & $  49.37 \pm  0.30$ &  6316 &   0.89 \\
   3241299 & $ 1117.6 \pm  3.0$ & $  57.80 \pm  0.18$ &  6316 &   1.10 &   8737094 & $  887.5 \pm  2.0$ & $  45.72 \pm  0.18$ &  6541 &   1.08 \\
   3850086 & $  989.9 \pm  2.5$ & $  52.09 \pm  0.58$ &  6461 &   0.92 &   8801316 & $ 1111.4 \pm  2.4$ & $  59.37 \pm  0.11$ &  6489 &   0.93 \\
   3852594 & $ 1006.4 \pm  2.2$ & $  53.64 \pm  0.21$ &  6417 &   0.85 &   8806223 & $ 1729.0 \pm  2.9$ & $  83.09 \pm  0.28$ &  6480 &   0.62 \\
   3936993 & $ 1195.3 \pm  4.4$ & $  61.89 \pm  1.33$ &  6225 &   0.73 &   8914779 & $ 1492.6 \pm  4.1$ & $  74.05 \pm  0.59$ &  6258 &   0.64 \\
   4484238 & $ 2080.2 \pm  2.1$ & $  92.80 \pm  0.07$ &  6043 &   0.56 &   9109988 & $ 2255.5 \pm  5.6$ & $  98.22 \pm  0.23$ &  6127 &   0.44 \\
   5105070 & $ 1154.8 \pm  2.2$ & $  61.93 \pm  0.47$ &  6090 &   0.62 &   9209245 & $ 1092.0 \pm  3.4$ & $  54.61 \pm  0.31$ &  6235 &   0.77 \\
   5183581 & $ 1017.8 \pm  1.7$ & $  53.81 \pm  0.19$ &  6387 &   0.87 &   9221678 & $ 1746.4 \pm  5.1$ & $  80.81 \pm  0.75$ &  6281 &   0.62 \\
   5597743 & $ 1060.7 \pm  3.3$ & $  55.07 \pm  0.59$ &  6014 &   0.84 &   9412514 & $  823.3 \pm  2.4$ & $  47.93 \pm  0.19$ &  6075 &   0.92 \\
   5696625 & $  704.9 \pm  2.7$ & $  39.29 \pm  0.18$ &  6402 &   1.48 &   9426660 & $ 1491.9 \pm  2.5$ & $  75.43 \pm  0.39$ &  6157 &   0.54 \\
   5771915 & $ 1420.6 \pm  5.3$ & $  70.55 \pm  0.74$ &  6230 &   0.68 &   9529969 & $ 1581.8 \pm  3.2$ & $  73.75 \pm  1.50$ &  6021 &   0.48 \\
   5791521 & $  669.0 \pm  2.7$ & $  34.45 \pm  0.26$ &  6398 &   1.35 &   9715099 & $  741.9 \pm  1.1$ & $  40.93 \pm  0.17$ &  6217 &   1.04 \\
   5856836 & $ 2104.9 \pm  5.0$ & $  91.87 \pm  0.59$ &  6233 &   0.48 &   9837454 & $  782.9 \pm  2.0$ & $  46.62 \pm  0.29$ &  5912 &   0.78 \\
   5871558 & $  862.6 \pm  2.4$ & $  47.23 \pm  0.28$ &  6239 &   0.88 &   9898385 & $ 1269.3 \pm  5.4$ & $  69.58 \pm  0.35$ &  6257 &   0.75 \\
   5905822 & $ 1686.0 \pm  1.7$ & $  82.86 \pm  0.10$ &  6057 &   0.50 &   9959494 & $ 1856.8 \pm  3.2$ & $  88.99 \pm  0.17$ &  6638 &   0.51 \\
   5982353 & $  940.9 \pm  3.3$ & $  47.77 \pm  0.16$ &  6208 &   1.04 &  10010623 & $  952.3 \pm  3.6$ & $  57.17 \pm  0.95$ &  6532 &   1.01 \\
   6048403 & $ 1039.9 \pm  2.0$ & $  53.58 \pm  0.15$ &  6444 &   0.89 &  10208303 & $ 1254.4 \pm  2.0$ & $  63.35 \pm  0.43$ &  6303 &   0.84 \\
   6062024 & $ 1060.7 \pm  2.2$ & $  57.59 \pm  0.17$ &  6166 &   0.81 &  10340511 & $ 1957.4 \pm  2.9$ & $  91.73 \pm  0.21$ &  5822 &   0.36 \\
   6268607 & $  789.7 \pm  3.3$ & $  41.71 \pm  0.63$ &  6208 &   0.95 &  10448382 & $  825.8 \pm  1.8$ & $  45.79 \pm  0.23$ &  5969 &   0.94 \\
   6359801 & $ 1309.2 \pm  6.8$ & $  64.85 \pm  1.11$ &  6463 &   0.78 &  10514274 & $ 1098.6 \pm  3.5$ & $  53.47 \pm  0.41$ &  6190 &   0.84 \\
   6761569 & $ 1057.4 \pm  1.3$ & $  58.27 \pm  0.34$ &  6271 &   0.83 &  10557075 & $ 1794.6 \pm  2.7$ & $  80.11 \pm  0.16$ &  5825 &   0.52 \\
   6777146 & $ 2090.3 \pm  4.9$ & $  92.55 \pm  0.66$ &  6103 &   0.47 &  10775748 & $ 1195.6 \pm  1.6$ & $  61.25 \pm  0.29$ &  6393 &   0.82 \\
   6784857 & $ 1241.9 \pm  1.8$ & $  63.70 \pm  0.39$ &  6172 &   0.71 &  10813660 & $ 1846.0 \pm  3.0$ & $  84.82 \pm  0.70$ &  6170 &   0.50 \\
   6881330 & $ 1790.1 \pm  3.8$ & $  84.03 \pm  0.26$ &  6092 &   0.49 &  10864215 & $  478.4 \pm  1.6$ & $  32.16 \pm  0.42$ &  6348 &   1.21 \\
   7205315 & $ 1345.9 \pm  5.3$ & $  66.59 \pm  0.50$ &  6051 &   0.75 &  10920182 & $ 1354.1 \pm  1.9$ & $  67.24 \pm  0.12$ &  5957 &   0.62 \\
   7215603 & $ 1790.2 \pm  1.9$ & $  83.65 \pm  0.26$ &  6230 &   0.54 &  11137841 & $  667.2 \pm  2.2$ & $  39.90 \pm  0.32$ &  6406 &   1.12 \\
   7272437 & $ 1229.9 \pm  2.2$ & $  63.62 \pm  0.39$ &  6078 &   0.70 &  11230757 & $ 1079.2 \pm  2.5$ & $  59.51 \pm  0.41$ &  6158 &   0.86 \\
   7418476 & $ 1241.8 \pm  4.0$ & $  63.95 \pm  0.28$ &  6725 &   0.74 &  11245074 & $  420.4 \pm  1.0$ & $  23.23 \pm  0.34$ &  6073 &   1.29 \\
   7439300 & $  809.6 \pm  1.6$ & $  42.15 \pm  0.17$ &  6121 &   1.03 &  11255615 & $ 1340.9 \pm  1.2$ & $  70.55 \pm  0.19$ &  6194 &   0.59 \\
   7500061 & $ 1307.3 \pm  2.9$ & $  63.50 \pm  0.15$ &  6265 &   0.73 &  11337566 & $  958.3 \pm  1.4$ & $  50.52 \pm  0.40$ &  6238 &   0.91 \\
   7669332 & $  503.9 \pm  1.8$ & $  43.87 \pm  1.08$ &  6183 &   1.13 &  11769801 & $  880.9 \pm  3.2$ & $  45.23 \pm  0.56$ &  6389 &   1.06 \\
   7833587 & $ 1302.9 \pm  3.0$ & $  64.91 \pm  0.22$ &  6236 &   0.72 &  11862497 & $ 2072.7 \pm  3.4$ & $  90.68 \pm  0.39$ &  6426 &   0.58 \\
   8165738 & $ 1622.5 \pm  4.3$ & $  82.40 \pm  1.25$ &  6331 &   0.53 &  11970698 & $ 1324.0 \pm  2.3$ & $  67.54 \pm  0.22$ &  5841 &   0.62 \\
   8243381 & $  879.8 \pm  3.6$ & $  50.26 \pm  1.51$ &  6101 &   0.81 &  12600459 & $ 2019.1 \pm  3.5$ & $  91.33 \pm  0.26$ &  6284 &   0.48 \\
\hline
\end{tabular}
\end{center}
\end{table*}

\begin{table*}
\begin{center}
\caption{Stars discussed in the text with possible solar-like oscillations.  The
columns are the same as in Table 1.}

\vspace{3mm}

\tiny

\begin{tabular}{rrrrrrrrrrrrr}
\hline
\multicolumn{1}{c}{KIC}                  & 
\multicolumn{1}{c}{$\nu_{\rm max}$}      & 
\multicolumn{1}{c}{$\Delta\nu$}          &
\multicolumn{1}{c}{$T_{\rm eff}$}         & 
\multicolumn{1}{c}{$\log\tfrac{L}{L_\odot}$}  & 
\multicolumn{1}{c}{KIC}                  & 
\multicolumn{1}{c}{$\nu_{\rm max}$}      & 
\multicolumn{1}{c}{$\Delta\nu$}          &
\multicolumn{1}{c}{$T_{\rm eff}$}         & 
\multicolumn{1}{c}{$\log\tfrac{L}{L_\odot}$}  \\
\hline
   3459226 & $  964.8 \pm  4.2$ & $  60.17 \pm  1.44$ &  7410 &   0.89 &  7868092 & $ 1709.9 \pm  7.6$ & $  73.31 \pm  1.39$ &  6726 &   1.16 \\
   5476495 & $  954.7 \pm  3.1$ & $  52.09 \pm  0.92$ &  7582 &   1.30 &  8258103 & $ 1866.5 \pm  3.3$ & $  80.83 \pm  0.40$ &  9082 &   1.28 \\
   5608334 & $ 2099.4 \pm 12.2$ & $  97.97 \pm  3.61$ &  6900 &   0.89 &  8264254 & $ 1534.6 \pm 11.0$ & $  72.33 \pm  3.65$ &  7174 &   0.96 \\
   6448112 & $ 2343.3 \pm  8.4$ & $ 122.80 \pm  2.25$ &  8498 &   2.06 &  9347095 & $  762.2 \pm  1.6$ & $  35.23 \pm  0.44$ &  7035 &   1.30 \\
   7212040 & $ 1313.9 \pm  4.4$ & $  57.29 \pm  0.60$ &  7710 &   0.94 &  9696853 & $ 1297.9 \pm  8.0$ & $  66.71 \pm  1.78$ &  7352 &   0.90 \\
\hline
\end{tabular}
\end{center}
\end{table*}

\label{lastpage}

\end{document}